\documentclass{article}
\usepackage{amsfonts}  
\usepackage{amsthm}
\usepackage{amsmath}
\evensidemargin=\oddsidemargin
\begin{document}
\title{The Hamiltonians of Linear Quantum Fields:\\
II.  Classically Positive Hamiltonians}
\author{Adam D. Helfer\\
Department of Mathematics\\ University of Missouri\\
Columbia, MO 65211, U.S.A.}
\maketitle
\begin{abstract}
For linear bose field theories, I show that if a classical Hamiltonian
function is strictly positive
in a suitable sense, the classical evolution must be conjugate, by a
symplectic motion, to a strongly continuous one-parameter orthogonal
group.  This can be viewed as an infinite-dimensional analog of the
existence of action-angle coordinates.  

This result is used to show that there is an intimate connection
between unitarity of the quantum evolution and boundedness-below of
the quantum Hamiltonians.  (Recent work has shown that generically, the
Hamiltonian operators of quantum fields in curved space--time are not
bounded below and do not generate unitary evolutions.)  Precisely, 
whenever the quantum Hamiltonians exist as self-adjoint operators,
they are bounded below.  Lower bounds on the
normal--ordered quantum Hamiltonian operators are computed.

Finally, it is shown that there is a broad class of ``quantum
inequalities:'' it $f^{ab}$ is a smooth compactly-supported
future-directed test function, then the operator $\int
f^{ab}{\widehat{T}}_{ab}\, d {\rm vol}$ is bounded below.
\end{abstract}
\newtheorem*{thmm}{Theorem}
\newtheorem*{corr}{Corollary}
\newtheorem{proposition}{Proposition}
\newtheorem{definition}{Definition}
\newtheorem{corollary}{Corollary}
\newtheorem{theorem}{Theorem}
\newtheorem{lemma}{Lemma}

\def\trace{\mathop{\rm tr}}
\def\tr{\mathop{\rm tr}}
\def\sech{\mathop{\rm sech}}
\def\sgn{\mathop{\rm sgn}}
\def\rSp{{\rm Sp}_{\rm rest}}
\def\Sp{{\rm Sp}}
\def\ix{{\rm i}}
\def\d{d}       
\def\e{e}       
\def\cA{{\mathcal A}}
\def\C{{\mathbb C}}
\def\R{{\mathbb R}}
\def\A{{\mathbb A}}
\def\Z{{\widehat Z}}
\def\wp{{\widehat\phi}}
\def\wH{{\widehat H}}
\def\H{{\wH}}
\def\X{{\hat X}}
\def\wpsi{{\hat\psi}}
\def\DA{{{\cal D}(A)}}

\section{Introduction}

This is the second of a three-part series analyzing the Hamiltonians of
linear quantum fields.  A general introduction will be found in the
first paper, Helfer (1999), hereafter called Part I.

When the evolution of the field operators in a linear quantum field
theory does not preserve their decomposition into creation and
annihilation parts, the analysis of the theory can be difficult
and is not yet wholly understood.  This situation can arise for several
reasons:

\begin{itemize}

\item
It occurs naturally in the presence of time-dependent
external potentials; in particular, it 
is the generic situation for quantum fields in curved
space--time.  

\item
It occurs when several
linear fields are linearly coupled, for instance, in models of
the quantum electromagnetic field in dispersive media.  

\item
It occurs when the evolution is not a perfect symmetry.  This happens,
for example, when the corresponding operator is an inhomogeneous
spatial or temporal average of the stress--energy.  Such operators are
the subject of the {\em quantum inequalities}, which bound the
persistence of negative energy densities, and are important in quantum
measurement issues.

\end{itemize}

In such situations, phenomena which are at least na\"\i{}vely very
pathological can occur.
In the case of quantum fields in
curved space--time, it has been shown that the quantum Hamiltonian
operators have only a restricted existence, being defined only as
quadratic forms.  (That is, there is a dense family of states for which
the expectations $\langle\Psi |{\widehat H}|\Psi\rangle$ are defined,
but there are no known non-zero states for which ${\widehat
H}|\Psi\rangle$ exists as an element in the physical Hilbert space.)
The quantum Hamiltonians' expectations are {\em unbounded below}, and
the corresponding evolutions are {\em not unitarily implementable}.
This means that the algebra of field operators does not evolve by
unitary motions.  This is distinct from the evolution of the state
vectors, which is unitary (and, in the usual ``relativistic
Heisenberg'' picture, trivial except for reductions).  Any sort of
non-unitarity in quantum theory should be taken seriously, and the
physical significance of that discovered recently is not yet clear.  One
of the main aims of these papers is to get a firm enough mathematical
control on the phenomenon that progress on a physical understanding
will be possible.

It may be helpful to comment on how these issues are related to the ``abstract
algebraic approach.''  In general, ``algebraic approaches'' seek to formulate
quantum field theory, as much as possible, in terms of the algebras of
observables.  They seek to avoid, or treat as derived concepts, the realization
of those observables as operators on Hilbert spaces.  

While such approaches have demonstrable power, the construction of a physical
representation is necessary to fully define the quantum theory, at least in a
conventional sense.  For example, the set of allowable 
$n$-point functions (with
the correct asymptotics) depends on the physical choice of
representation, not just on the algebra of observables.  In general, while some
of the physical content is specified in the algebraic structure alone, the full
physical content depends on knowing the representation.  For instance,
Hawking's (1974, 1975) prediction of black-hole evaporation relies strongly on
the choice of representation.  (See Helfer 2003 for a critique of this
prediction.)

The unboundedness-below and failure of unitary implementability are
problems which are much more apparent in an approach based on the
construction of representations than in an abstract algebraic
approach.  Since the aim of these papers is to develop a framework in
which the significances of these issues can be assessed, the approach
here is very much based on the study of representations.  

Which sort of approach will ultimately --- when the significances of
these issues {\em are } understood --- be most appropriate remains to
be seen.  Should the apparent pathologies turn out to be simply
mathematical fine points not of real physical consequence, an
algebraic approach might well be the most suitable.  On the other
hand, unitarity is such a basic concept of quantum theory, 
and boundedness-below of
energy such a fundamental concern in any physical theory, that it may
well turn out that these issues {\em are } important and are best
understood within a representation.

In Part I of this series, I determined under what conditions an
infinitesimal symmetry of the classical phase space gave rise to a
self-adjoint quantum operator.  In that paper, no special properties
of the symmetry were used.  In the present paper, I specialize to those
symmetries corresponding to positive classical Hamiltonian
functions.  These arise in particular for those which are energy
operators, in the sense that they correspond to evolution forward in
time.

\paragraph{Structure Theory}

The first aim of this paper is to classify the different possible
structures such classical Hamiltonians might have.  This will then be
used to analyze their quantizations.

Theorem~\ref{gayatri}, then,
is a classification of the possible
classical Hamiltonians.  
Essentially, it states that a classically positive
Hamiltonian function generates a family of motions which is similar, by
a canonical transformation, to a one-parameter orthogonal
group. (However, the theorem as stated and proved here requires
an additional technical hypothesis,
which is believed not to be necessary.  See the
discussion preceding the theorem.)
This comes about because even in
this infinite-dimensional context there is something like a
compactness of the constant-energy surfaces in the classical
phase space.  This is a delicate and remarkable result, which is
presumably of interest in the theory of infinite-dimensional dynamical
systems.  It implies an infinite-dimensional analog of the existence
of action-angle variables, for example.  It also implies that the
classical evolution remains uniformly bounded in time.
This extends
the structure theory for classical Hamiltonian field theories that was
developed earlier (see Chernoff and Marsden 1974).

\paragraph{Self-Adjointness and Boundedness-Below}

Using the classical structure theorem to get a handle on the quantum
theory, we find (Theorem~\ref{binswanger}) that a classically positive
Hamiltonian is self-adjointly implementable iff the similarity
effecting its transformation to a generator of orthogonal motions
corresponds to a restricted Bogoliubov transformation, and, in this
case, the quantum Hamiltonian must be bounded below.\footnote{Recall
that the restricted Bogoliubov transformations are those which lead to
unitarily equivalent quantizations.}  Thus one has a
very strong connection between self-adjointness and
boundedness-below, for energy operators.  It should be emphasized that
these results make no presupposition about what renormalization
prescription is to be used.

There is an old folk-theorem in quantum field theory:  ``A Hamiltonian
determines its quantization,''  meaning that a formal expression for a
Hamiltonian should have a unique (modulo c-numbers) implementation as
an operator.  Theorem~\ref{binswanger} can be viewed as allowing one to
make this statement precise, for linear field theories.  It shows that
classically positive Hamiltonians have certain mathematically allowed
quantizations, which may or may not be physically acceptable, according
to whether the Bogoliubov transformation is restricted.  One might
think that the correct interpretation of the present results is simply
that, when the Bogoliubov transformations turn out not to be
restricted, one picked the ``wrong'' original set of canonical
variables, and one should choose another, leading to an allowed
quantization.  However, at least for the case of quantum fields in
curved space--time, this does {\em not} seem to be the correct
interpretation.  There, the acceptable choices of canonical variables
for quantization are determined by the ``Hadamard'' condition, and in
general the Hamiltonians are not compatible with this.

In other words, the physical considerations leading to a choice of
canonical variables, and those leading to the choice of Hamiltonian,
conflict in that the Hamiltonian {\em cannot } be self-adjointly
realized.  It should also be mentioned that even if one chooses a
mathematically allowed quantization giving rise to a self-adjoint
Hamiltonian on one hypersurface in space--time, one would need
inequivalent representations for the Hamiltonians at nearby
hypersurfaces.

\paragraph{Normal-Ordering and Lower Bounds}

The results described so far are general statements about when a
classical Hamiltonian is self-adjointly implementable at the quantum
level, without presupposing a specific renormalization scheme.  One
would like to know what relation these results bear to standard
renormalization theory.  A choice of renormalization prescription is
necessary, too, to move beyond the statement that the Hamiltonians are
bounded below and be able to speak of their lower bounds (since
renormalization in particular determines the c-number contribution to
the Hamiltonians).

Almost universally for linear fields, normal-ordering
(or an equivalent prescription, like point-splitting) is used as the
prescription.  We show that, for classically positive operators, normal
ordering may not suffice to renormalize the Hamiltonian, and that
finiteness of the normally-ordered ground-state energy is equivalent
to the existence of the normally-ordered Hamiltonian as a
self-adjoint operator (Theorems~\ref{albert} ands \ref{liese}).  We
also compute some lower bounds for the normally-ordered ground state
energy (Theorem~\ref{jason}).

\paragraph{Quantum Inequalities}

For some time now, it has been conjectured that suitably
temporally averaged measures of the energy density operator for
quantum fields in curved space--time should be bounded below.  Such
bounds are known as {\em quantum inequalities}, following pioneering
work of Ford (1978).  We establish the existence (but not the precise
numerical bounds) of a wide class of these:  for any smooth,
compactly-supported test function $f^{ab}$ which is {\em
future-directed} in the sense that it is at each point a symmetrized
product $t^au^b+u^at^b$ of future-directed vectors, the operator
\begin{equation}
\int {\widehat T}_{ab} f^{ab}\, \d {\rm vol}
\end{equation}
is bounded below (Theorem~\ref{greenspan}).  The significance of this
result is discussed more fully at the beginning of section~\ref{vort}.

Unfortunately, in order to establish these inequalities, we cannot make
direct use of the earlier structure results.  This is because the
Hamiltonians corresponding to averages of components of the
stress--energy over compact regions of space--time, do not satisfy the
strong form of classical positivity needed for those theorems.  The
quantum inequalities are proved from a partial result on the structure
theory for these ``weakly positive'' classical Hamiltonians,
Theorem~\ref{krypton}.  

The organization of the paper is this.  Section 2 contains some
preliminaries.  Section 3 goes over the main structure of classically
positive Hamiltonians; in Section 4, these results are applied to
quantum field theory.  Section 5 works out the connection with normal
ordering, and Section 6 the structure necessary to establish the
existence of the quantum inequalities.  The last section contains some
discussion.

{\em Summary of Notation.}  Here is a summary of the notation used. 
Unfortunately, there are quite a few things denoted conventionally by
similar symbols.

$H$ is the space of solutions of the classical field equations, a real
separable Hilbertable space equipped with a symplectic form $\omega$.

$H_\C$ is the space $H$ equipped with the complex structure defined by
$J$, and so made into a complex Hilbert space.

${\cal H}$ is the Hilbert space on which the representation acts.

$\|\cdot\| _{\rm op}$ is the operator norm.

$\|\cdot\| _{\rm HS}$ is the Hilbert--Schmidt norm.

$\A$ is the field algebra.

$A$ is the Hamiltonian vector field on the space of classical solutions.

$\cA$ is the Lie adjoint of $A$, that is, the derivative of conjugation 
by $g(t)=\e ^{tA}$.

{\em Note.}  Since $H$ is not canonically a Hilbert space, I will
generally emphasize the dependence of properties on the choice of norm,
that is, of $J$.  Thus one has $J$-linear, rather than complex-linear,
transformations.  Similarly, there are $J$-symmetric, $J$-orthogonal,
etc., transformations.

\section{Preliminaries}

Throughout, we shall let $H_\C$ be a complex infinite-dimensional
separable Hilbert space.  The complex inner product on $H_\C$ will be
denoted $\langle\cdot ,\cdot\rangle$.  We shall let $H$ be the underlying
real Hilbert space.  Then we write $J:H\to H$ for the real-linear map
given by $v\mapsto \ix v$, and
\begin{eqnarray}
 (v,w)&=&\Re\langle v,w\rangle\\
     \omega (v,w)&=&\Im \langle v,w\rangle
\end{eqnarray}
Then $(\cdot ,\cdot )$ is the canonical real inner product on $H$ and
$\omega$ is a symplectic form on $H$ which is non-degenerate in that it
defines isomorphisms from $H$ to its dual.  Note that
\begin{equation} (v,w)=\omega (v,Jw )\, .\end{equation}
Thus any two of $\omega$, $J$ and $(\cdot ,\cdot )$ determine the third.

Throughout, the {\em real } adjoint of a real-linear operator (perhaps
only densely defined) $L$ will be denoted $L^*$.  Thus the defining
relation is $(v,L^*w)=(Lv,w)$ with domain $D(L^*)=\{ w\in H\mid
(v,L^*w)=(Lv,w)$ for some $L^*w$ for all $v\in D(L)\}$.  

\begin{definition}
The {\em symplectic group} of $H$ is
\[ \Sp (H)=\{ g:H\to H\mid g\mbox{ is linear, continuous, and preserves }
  \omega\}\, .
\]
Its elements are the {\em symplectomorphisms.}
\end{definition}

The symplectic group does not depend on the real inner product on $H$
(or on the complex structure); it depends only on $\omega$ and the
structure of $H$ as a Hilbertable space.  It has naturally the structure
of a Banach group, using the operator norm to define the topology.

\begin{definition}
The {\em restricted symplectic
group } of $H_\C$ is 
\[\rSp (H_\C )=\{g\in \Sp (H)\mid
  g^{-1}Jg-g\mbox{ is Hilbert--Schmidt}\}\, .\]
\end{definition}

We recall that a {\em strongly continuous} one-parameter subgroup of
$\Sp (H)$ is a one-parameter subgroup $t\mapsto g(t)$ such that, for
each $v\in H$, the map $t\mapsto g(t)v$ is continuous.  (In general,
one can also consider {\em semigroups}, defined for $t\geq 0$, but as
every symplectomorphism is invertible, in our case every semigroup
extends to a group, which is strongly continuous iff the semigroup is.)
According to the Hille--Yoshida--Phillips Theorem, such groups have the
form $g(t)=\e ^{tA}$, where $A$ is a densely--defined operator on $H$
(with certain spectral properties), and $\| g(t)\| _{\rm op}\leq M\e
^{\beta |t|}$ for some $M,\beta \geq 0$.  
The spectrum of $A$ is confined to the
strip $| \Re\lambda |\leq\beta$.

\section{Classically Positive Hamiltonians}

The analysis so far has been concerned with general symmetries of the
phase space.  In the case of time evolution, there are important
additional properties.  The most fundamental of these is that, in the
classical context, the energy cannot be negative.  Indeed, this fact
plays a key role in establishing the existence and stability of
temporal evolution from initial data.  In this section, we shall
investigate this extra structure.

A key result is that when the energy function is positive, the
evolution must be conjugate to a unitary group.  This is quite
remarkable, even in the case of finite dimensions, since the
eigenvalues of a general Hamiltonian vector may be complex.  On the
other hand, the result is (in finite dimensions) essentially an
extension of the proof of the existence of action--angle variables.

The argument in finite dimensions is this.  Let $A$ be the generator of
a one-parameter group of symplectic transformations, and suppose its
energy function $(1/2)\omega (v,Av)$ is a positive-definite quadratic
form.  Since evolution by $g(t)=\e ^{tA}$ preserves this form, we see
that $g(t)v$ remains bounded for all $t$, for any $v$.  This means that
the eigenvalues of $A$ must be purely imaginary, and that its Jordan
form (over the complex) must be purely diagonal.  Thus $A$ must be
conjugate to an anti-Hermitian matrix.

In infinite dimensions, the result is made more difficult for several
reasons.  In the first place, since the operator $A$ is unbounded, the
form $\omega (v,Av)$ is not defined everywhere.  This means that for a
dense family of $v$'s, the form has the value $+\infty$, and the fact
that this value remains constant in $t$ does not allow us to conclude
that the orbits $g(t)v$ remain bounded.  Also, it is not known {\it a
priori } that $A$ (or even $g(t)$) has anything like a Jordan normal
form.  (Indeed, the required property, known as ``spectrality,'' is in
general a very delicate thing to establish.  An example of a
Hamiltonian with $A$ nonspectral was given in Paper I.) In fact, our
argument turns on a recently established hyperfunctional analog of a
theorem of Bochner, and it is possible that the lack of adequate
analytic tools prevented an earlier proof.

\begin{definition}
The generator $A$ of a strongly-continuous one-parameter subgroup of
the symplectic group is called {\em classically positive } if the form
$(1/2)\omega (v,Av)\geq c\| v\| ^2$ for some $c>0$.
\end{definition}

The requirement that $c$ be strictly positive will be used essentially
in what follows.  Its effect is to rule out certain potential infrared
problems.  (The analysis could be modified to accommodate a finite number
of zero modes of $A$, however.)

The next theorem is one of our main results.  It asserts that (with one
technical proviso) classically positive Hamiltonians are in fact
similar, by bounded symplectomorphisms, to the generators of orthogonal
motions on phase space.  Thus this is a general structure theorem, which
can be thought of as an analog of the statement that action--angle
variables exist for linear systems with positive Hamiltonians and
finitely many degrees of freedom.

As mentioned above, the theorem contains a technical proviso, which is
that $A_-$, the $J$-antilinear part of $A$, be bounded.  (So the
theorem would apply to any $A$ such that an some positive complex
structure $J$ could be found for which $A_-$ is bounded.)  This
condition is verified in all examples known to me, and holds in
particular for quantum fields in curved space--time (Helfer 1996).
Still, it would be be more satisfying to remove this hypothesis, and I
believe this can be done.  However, the arguments if unbounded $A_-$
are allowed are much more technically complicated, will be pursued
elsewhere.

\begin{proposition}\label{phyllis}
Let $A$ be classically positive.  Then its spectrum lies on the
imaginary axis.
\end{proposition}

\begin{proof}
Let ${\cal D}(A)$ be the domain of $A$.  The spectrum of $A$ is the set
of points $\lambda$ at which the map $\lambda -A:{\cal D}(A)\to H$ is
not invertible.  More precisely, since $H$ is not canonically a complex
vector space, we work with the complexification.  This will be done in
the usual way, without introducing unnecessary notations for
complexifications.  Then $\omega (\overline{v} ,Av)$ is a Hermitian
form bounded away from zero (as a form).

Suppose $\lambda -A$ is not one-to-one.  Then $A$ has an eigenvector
$v$ with eigenvalue $\lambda$.  Then
\begin{eqnarray*}
 \| \e ^{\lambda t}v\| ^2&=& \| g(t)v\| ^2\\
  &\leq& (2c)^{-1}\omega\bigl( g(t){\overline v},Ag(t)v\bigr)\\
  &=& (2c)^{-1}\omega (\overline{v} ,Av)\, .
\end{eqnarray*}
This can hold  for all $t$ only if $\lambda$ is purely imaginary (or
zero).

Now suppose that $\lambda -A$ is not onto, and its image lies in some
hyperplane $\{ x \mid (v,x)=0\}$.  This means that $v$ is an
eigenvector of $A^T$ with eigenvalues $\lambda$.  Now $A^T$ is the
generator of $g^T(t)=-Jg(-t)J$.  The domain of $A^T$ is $J{\cal D}(A)$,
and the energy function is $(1/2)\omega (y,A^Ty)=(1/2)\omega
(y,JAJy)=-(1/2)\omega \bigl( (Jy),A(Jy)\bigr)$.  As in the previous
paragraph, we find
\[ \| \e ^{\lambda t}v\| ^2\leq -(2c)^{-1}
  \omega (\overline{v} ,A^Tv)\, ,\]
forcing the real part of $\lambda$ to vanish.

We now take up the more delicate case, where $\lambda -A$ is
one-to-one but not onto, but its image is dense.  In this case, we may
find a sequence $v_n$ of unit vectors in $H$ which are elements of 
${\cal D}(A)$ such that $\| v_n\|
+\| Av_n\| =1+\| Av_n\|$ 
is bounded and $(\lambda -A)v_n$ tends to zero.  We have
\[ \bigl( g(t)-\e ^{\lambda t}\bigr) v=\e ^{\lambda t}\int _0^t
g(u)\e ^{-\lambda u}\d u\, (A-\lambda )v\]
for any $v\in {\cal D}(A)$.  Thus
\[ \| \bigl( g(t)-\e ^{\lambda t}\bigr) v\| \leq  M
  \bigl( \e ^{\beta t}-\e ^{\Re \lambda t}\bigr) (\beta -\Re\lambda )^{-1}
    \| (A-\lambda )v\| \]
where we have taken $t\geq 0$ for simplicity (and we have used the
Hille--Yoshida--Phillips bound $\| g(t)\| _{\rm op}\leq M\e
^{\beta |t|}$).

Now, suppose $\Re\lambda >0$.  We may choose $T>0$ so that  $\e ^{\lambda
T}$ is as large as desired.  For such a $T$, and any $\epsilon >0$, we
may choose $n$ large enough so that $\| \bigl( g(t)-\e ^{\lambda t}\bigr)
v_n\| <\epsilon$ for all $0\leq t\leq T$.  In this case we will have 
\[ \| \e ^{\lambda t}v_n\| ^2\leq\bigl( \| g(t)v_n\| +\epsilon\bigr) ^2
  \leq\bigl( \sqrt{(2c)^{-1}\omega (\overline{v_n},Av_n)} 
  +\epsilon\bigr) ^2\, .
\]
However, this is a contradiction, for the right-hand side is uniformly bounded,
and $\| e^{\lambda t}v_n\| \allowbreak
=e^{\Re\lambda t}$ can be made as large as desired.
Thus $\Re\lambda \not{>} 0$.

Consideration of $t<0$ similarly rules out the case $\Re\lambda <0$, and
so we must have $\Re\lambda =0$.  

\end{proof}

\begin{theorem}\label{gayatri}
Let $A$ be classically positive, and suppose its $J$-antilinear part
$A_-$ is bounded.
Then there is a positive-definite
bounded $J$-symmetric symplectomorphism $\gamma$, and a
$J$-real-anti-self-adjoint $J$-linear closed (possibly unbounded)
operator $\sigma$, such that
\[ g(t) =\gamma^{-1}\e ^{\sigma t}\gamma\, .\]
There is a $J$-orthogonal $J$-invariant
projection-valued measure $\d F(\theta )$ supported on $[\theta
_0,\infty )$ for some $\theta _0>0$ such that
\[ \e ^{\sigma t}=\int _\R \e ^{t\theta J}\d F(\theta )\, .\]
\end{theorem}

\begin{proof}
The idea will be to define a positive complex structure $J_A$ relative
to which $g(t)$ is orthogonal.  The conclusions will follow almost
directly from this.  At a formal level, one has
\[ J_A=\pi ^{-1}\int _{-\infty}^\infty (\lambda -A)^{-1}\d\lambda\,
.\]
However, the sense in which this integral converges as
$\lambda\to\pm\infty$ needs to be made precise.
Even if $A$ were known to be bounded, the existence of this integral would be a
bit delicate.  In the present case, there are a number of technicalities, which
arise because we need to gradually establish enough properties of $J_A$ to show
that it exists as a bounded operator.  Once this is done, the remainder of the
proof will be routine algebraic computations.

We begin by showing that $J_A$ exists 
(as a potentially unbounded operator) on the
dense domain $D(A)$.  

Thr integral for $J_A$ converges strongly on $D(A)$ in
the sense of a Cauchy principal value.  
To see this, first note that since $\lambda ^{-1}$
converges near infinity as a Cauchy principal value, it is enough to
show that
\[ {\rm P}\, \int _{|\lambda |\geq \lambda _0}
  \left[ (\lambda -A)^{-1} \lambda ^{-1}\right] v\d\lambda\]
converges, where ``P'' indicates the principal value and $v\in D(A)$.
The integrand can be re-written:
\[
 \left( (\lambda -A)^{-1} -\lambda ^{-1}\right) v
  = \lambda ^{-1} (\lambda -A)^{-1} Av\, .\]
On the other hand, the   Hille--Yoshida--Phillips estimate $\|
g(t)\|_{\rm op}\leq M\e ^{\beta |t|}$ implies 
\[
\| (\lambda -A)^{-1} \|_{\rm op}\leq M(\Re\lambda -\beta )^{-1}
\]
for sufficiently large $\Re\lambda$ (and similarly for negative
$\Re\lambda$).
From this estimate, we have, for $v\in D(A)$,
\[ \left(  (\lambda -A)^{-1} -\lambda ^{-1}\right) v
 =\lambda ^{-1}(\lambda -A)^{-1} Av\]
of order $O(\lambda ^{-2})$ as $\Re\lambda\to \infty$. This is integrable at
infinity, and so $J_A v$ exists for $v\in D(A)$.

It is straightforward to verify that $J_A^2=-1$ in a suitable sense,
namely on $D(A^2)$.  This is a singular-integral computation using the
definition of $J_A$. 
We first re-write the principal-value integral:
\[ J_A=\pi ^{-1}\int _0^\infty\left[ (\lambda -A)^{-1} +(-\lambda
-A)^{-1}\right]\, \d\lambda\]
(understood strongly on $D(A)$).  Now we have
\[ J_A^2=\pi ^{-2}\int _0^\infty\d\lambda\int _0^\infty\d\mu
  \left[ (\lambda -A)^{-1} +(-\lambda
-A)^{-1}\right]
\left[ (\mu -A)^{-1} +(-\mu
-A)^{-1}\right]\]
(understood strongly on $D(A^2)$).
After some algebra, this can be re-written as
\[ J_A^2=\pi ^{-2}\int _0^\infty\d\lambda\int _0^\infty\d\mu
  \left[ (- \lambda ^2 +A^2)^{-1} - (-\mu ^2 -A ^2)^{-1}
    \right]  \frac{A^2}{\lambda ^2-\mu ^2}
\, .
\]
While the existence of this as a Riemann integral requires the cancellation of
$(-\lambda ^2 +A^2)^{-1}$ against $(-\mu ^2 +A^2)^{-1}$ in order to
compensate for the singularity in the denominator $\lambda ^2-\mu ^2$ at
$\lambda =\mu$, we may break the integral into two if each is interpreted as a
principal value:
\[ 
\begin{split}
J_A^2=&\pi ^{-2}\,{\rm P}\,\int _0^\infty\d\lambda\int _0^\infty\d\mu
  (- \lambda ^2 +A^2)^{-1} \frac{A^2}{\lambda ^2-\mu ^2}\\
  &- \pi ^{-2}\,{\rm P}\,\int _0^\infty\d\lambda\int _0^\infty\d\mu
 (-\mu ^2 -A ^2)^{-1}\frac{A^2}{\lambda ^2-\mu ^2}
\, .
\end{split}
\]
(This will of course be compatible with definition of the Riemann integral,
since it simply corresponds to a particular way of forming the limit which is
the integral.) 
If we do this, then we can use the distributional identity
\[ \int _0^\infty (\lambda ^2-\mu ^2) ^{-1}\d\mu =-(\pi ^2 /2)\delta (\lambda
)\]
on the first integral, and its counterpart for integration
over $\lambda$ on the second, to obtain
\[ J_A^2=-1\]
immediately, strongly on $D(A^2)$.

In the next stage of the analysis, we shall want to consider the
$J$-linear and antilinear portions of $J_A$.  Formally, these are
given by $(J_A)_\pm =(1/2)(J_A \mp JJ_A J)$.  However, in order to
make sense of this, we must show that $J_A$ and $JJ_AJ$ have a common
dense domain.

We shall show that the $J$-invariant dense set
$D(A^2)+JD(A^2)\subset D(A)$; then $(J_A)_\pm$ are naturally
defined on this domain.  (Here $D(A^2)=\{ v\in H\mid v=A^{-2}w$
for some $w\in H\}$.)  It is clear that $D(A^2)\subset D(A)$; we must
show that $JD(A^2)\subset D(A)$.  So let $v=JA^{-2}w\in JD(A^2)$.
Then
\begin{eqnarray*}
 v&=&A^{-1}(AJ)A^{-2}w\\
  &=& A^{-1} (JA-2JA_- )A^{-2}w\\
  &=& A^{-1}(JA^{-1}-2JA_-A^{-2})w\, ,
\end{eqnarray*}
which is an element of $D(A)$. The same sort of argument shows
$D(A^n)+JD(A^n)\subset D(A^{n-1})$ for $n\geq 1$.

Now in fact $(J_A)_-$ exists as a bounded operator.  
To see this, we will show that the $J$-antilinear part $\left( (\lambda
-A)^{-1}\right) _-$ of the resolvent is $O(\lambda ^{-2})$ in operator norm.  
This will
be accomplished by estimating
\[ \left( (\lambda -A)^{-1}\right) _-=\int _0^\infty g_-(t)\e ^{-\lambda t}\, \d
t\, .\]
(Using the symmetry $t\mapsto -t$, $A\mapsto -A$, it is enough to consider
the case of positive $\Re\lambda$.)
In section 4 of paper I, we considered a quantity $L(t)=2g_-(t)Jg(t)$ 
and showed
\[ L(t) =\int _0^t G(u)(2A_-J)\, \d u\]
where
\[ G(u)Q=g(u)Qg(-u)\, .\]
Using this, we find that for $\Re\lambda$ sufficiently large
\begin{eqnarray*}
\int _0^\infty g_-(t)\e ^{-\lambda t}\, \d t
  &=& (-1/2)\int _0^\infty L(t)g(-t) \e ^{-\lambda t}\, \d t\\
  &=& (-1/2)\int _0^\infty L(t)(-\lambda -A)^{-1}\frac{d}{dt}
    g(-t) \e ^{-\lambda t}\, \d t\\
  &=& (1/2)\int _0^\infty \left( G(t)(2A_-J)\right)
     (-\lambda -A)^{-1}g(-t) \e ^{-\lambda t}\, \d t
     \, .
\end{eqnarray*}        
This Hille--Yoshida--Phillips estimates, and the boundedness of $A_-$, now imply
this is $O(\lambda ^{-2})$ at infinity in operator norm.     

We remark that the boundedness of $(J_A)_-$ implies that $J_A$ extends
naturally to exist on $D(A)+JD(A)$, for we can define $J_AJ=J(J_A)_+-J(J_A)-$. 
However, we shall see shortly that in fact $J_A$ is itself bounded.

We next note that $\omega (\cdot ,J_A\cdot )$ is a positive-definite
symmetric form on $D(A)$.  (This follows easily from the fact that $A$
is a generator of symplectomorphisms and that it is classically
positive.)  Similarly, conjugating by $J$, we have $\omega (\cdot
,-JJ_AJ\cdot )$ a positive-definite symmetric form on $JD(A)$ (or $D(A)$).
Thus $(J_A)_+$ also defines a positive-definite symmetric form, and
hence $-J(J_A)_+$ is a $J$-symmetric operator defining a
positive-definite form $(\cdot ,-J(J_A)_+\cdot )=\omega (\cdot
,(J_A)_+\cdot )$ on $D(A)$.  Thus $-J(J_A)_+$ has a canonical
extension to a positive self-adjoint operator on a dense domain in
$H$.  (We shall continue to denote this operator by $-J(J_A)_+$.)

We now re-write the equation $J_A^2=-1$ in terms of the self-adjoint
operators $-J(J_A)_\pm$.  We have
\[ \left( (J_A)_+ +(J_A)_-\right) ^2 =((J_A)_+)^2 +((J_A)_-)^2
    +(J_A)_+(J_A)_- +(J_A)_-(J_A)_+=-1\, ,
\]
and the $J$-linear and $J$-antilinear parts of this are
\[ ((J_A)_+)^2 +((J_A)_-)^2=-1\qquad\text{and}\qquad
  (J_A)_+(J_A)_- +(J_A)_-(J_A)_+=0\, .\]
Multiplying through by $(-J)^2$, we get
\[ (-J(J_A)_+)^2 -(-J(J_A)_-)^2=1\text{ and }
  (-J(J_A)_+)(-J(J_A)_-) -(-J(J_A)_-)(-J(J_A)_+)=0\, .\]
From the first equation and the boundedness of $-J(J_A)_-$, we see
that $-J(J_A)_+$ (and hence $J_A$) is bounded.  According to the
second, the operators $-J(J_A)_+$, $-J(J_A)_-$ commute.  Thus, again
using the first equation, we may find a $J$-symmetric, $J$-antilinear
operator $\Theta$ such that 
\[ -J(J_A)_-=\sinh 2\Theta \qquad\text{and}\qquad -J(J_A)_+=\cosh
2\Theta \, .\]
(The factor of two is for later convenience.)
It will be useful to rewrite these.  Note that
\[ (J_A)_-=J\sinh 2\Theta  =\sinh (2J\Theta )
\qquad\text{and}\qquad (J_A)_+=J\cosh 2\Theta =J\cosh (2J\Theta )\, ,\]
where the $J$-antilinearity of $\Theta $ has been used (note that
$(J\Theta )^2=+(\Theta )^2$). 

We now let 
\[\gamma =\cosh\Theta -J\sinh\Theta =\exp (-J\Theta )\, .\]
Since
\[ \omega (v,J\Theta w)=\omega (\Theta v,Jw)=-\omega (J\Theta v, w)\,
,\]
the operator $J\Theta$ is a generator of symplectomorphisms and
$\gamma$ is a symplectomorphism.  We also have
\begin{eqnarray*}
  \gamma ^{-1}J\gamma &=&\e ^{J\Theta}J\e ^{-J\Theta}\\
    &=&J\e ^{-J\Theta} \e ^{-J\Theta}\\
    &=&J\left( \cosh (2\Theta )-J\sinh (2\Theta )\right) \\
    &=&J_A\, .
\end{eqnarray*}

Now we shall show that $\gamma 
g(t)\gamma ^{-1}$ is $J$-orthogonal:
\begin{eqnarray*}
 \omega (\gamma g(t)\gamma ^{-1}v,J\gamma g(t)\gamma ^{-1}w)
  &=& \omega (v,\gamma g(-t)\gamma ^{-1}J\gamma g(t)\gamma ^{-1}w)\\
 &=& \omega (v,\gamma g(-t)J_Ag(t)\gamma ^{-1}w)\\
 &=&\omega (v,\gamma J_A\gamma ^{-1}w)\\ 
 &=&\omega (v,Jw)\, .
\end{eqnarray*}
(Here we have used the fact that, by construction, the operator $J_A$ is
invariant under conjugation by $g(t)$.)
Since $\gamma g(t)\gamma ^{-1}$ is $J$-orthogonal and a symplectomorphism,
it is $J$-linear.  We may thus set
\[ \sigma =\gamma ^{-1}A\gamma\, .\]
\end{proof}

\section{Application to Quantum Field Theory}

We now apply the structure theorems of the previous section to quantum
field theory.

\begin{theorem}\label{binswanger}
In order that a classically positive operator $A$ 
with bounded antilinear part generate a
one-parameter group of restricted symplectic motions, it is
necessary and sufficient that the operator $\gamma$ be a restricted
symplectomorphism.  In this case the corresponding Hamiltonian 
operator is bounded below.
\end{theorem}

\begin{proof}
The $J$-antilinear part of $g(t)$ is
\begin{eqnarray*}
\lefteqn{J\sinh\Theta \e ^{t\sigma}\cosh\Theta +\cosh\Theta \e ^{t\sigma}
  (-J\sinh\Theta )}\\
 &&=J\bigl( \sinh\Theta \e ^{t\sigma} \cosh\Theta
  -\cosh\Theta \e ^{t\sigma}\sinh\Theta\bigr)\, .
\end{eqnarray*}
Multiplying on the left by $-J\sech\Theta$ and on the right by
$\sech\Theta$ (both bounded operators with bounded inverses), we see
that the antilinear part of $g(t)$ is Hilbert--Schmidt iff
\[ \tanh\Theta \e ^{t\sigma} -\e ^{t\sigma}\tanh\Theta\]
is, or equivalently if
\[ \tanh\Theta -\e ^{t\sigma}\tanh\Theta \e ^{-t\sigma}\]
is.

The idea now will be to think of $\tanh\Theta$ as a vector in the space of
symmetric bounded operators, and consider the action of $\e ^{t\sigma}$
on this space by conjugation.  However, this space is not a Hilbert space,
and in order to take advantage of the spectral theory of operators on
Hilbert space, it is more convenient to regard $\tanh\Theta$ as a sort
of unbounded form on the Hilbert--Schmidt operators.

Let $V$ be the space of Hilbert--Schmidt $J$-antilinear,
$J$-symmetric endomorphisms of $H$.  This space is naturally a
complex Hilbert space, with complex structure given by $L\mapsto JL$
and inner product
\[\langle\!\langle M,L \rangle\!\rangle =\trace (ML)+\ix\trace (LJM)\, .\]
Conjugation by $\exp \sigma t$ is a strongly
continuous unitary map on this space.  We may apply the usual spectral
theory of one-parameter unitary groups on Hilbert space to this.

In fact, we can work out the spectral resolution explicitly in terms of
that for $\exp \sigma t$.  For we have
\begin{eqnarray*}
  \e ^{\sigma t}L\e ^{-\sigma t}
   &=&\int \e ^{\theta t J}\d F(\theta )L\e ^{-\phi tJ}\d F(\phi )\\
  &=&\int \e ^{(\theta +\phi ) t J}\d F(\theta )L\d F(\phi )\\
  &=&\int \e ^{\xi t J}\d {\cal E}(\xi ) L
\end{eqnarray*}
where we have defined $\d {\cal E}(\xi ) L=\int _{(\theta ,\phi )\mid
\theta +\phi =\xi }\d F(\theta )L\d F(\phi )$.  One can check that $\d
{\cal E}$ is a projection-valued measure on $V$, and the equation above
provides the spectral resolution of conjugation by $\e ^{\sigma t}$.
Note that since $\d F(\theta )$ is
supported for $\theta\geq \theta _0>0$, the measure $\d {\cal E}(\xi )$
is supported for $\xi\geq 2\theta _0>0$.  
(This may be counterintuitive, as one thinks of the
generator of a one-parameter group of conjugations as having
eigenvalues which are {\em differences} of the eigenvalues of the
generator of the original group.  However, in the present case there is
a very curious interaction between the fact that the spectrum is
imaginary and the antilinearity of the elements of $V$.  This is in some
sense the central point of the proof.)

The above analysis does not quite apply directly to $\tanh\Theta$ or
$\Theta$, since $\Theta$ may not be Hilbert--Schmidt.  However, the
operator $\Theta$ is bounded, and so can be regarded as a linear
functional on the space $V_0$ of trace-class elements of $V$.  The
space $V_0$ is dense in $V$ and invariant under conjugation by
$\e ^{\sigma t}$, and so the spectral resolution derived above for this
conjugation can be applied, by duality, to $\Theta$, and similarly to
$\tanh\Theta$.

That $g_-(t)$ be Hilbert--Schmidt is thus equivalent to requiring
\[ \int _{[2\theta _0 ,\infty)} \left( \e ^{\xi t J}
   -1\right) \d {\cal E}(\xi
)\tanh\Theta\]
to be so.  Since $\d {\cal E}(\xi )$ resolves $V$ into the orthogonal
direct integral of Hilbert--Schmidt operators, the integral above,
restricted to any compact interval
of $\xi$-values, must be Hilbert--Schmidt.  This may be seen
to be equivalent to the requirement $\tanh\Theta$ be Hilbert--Schmidt by
elementary arguments.  And since $\Theta$ is $J$-symmetric, this is
equivalent to $\Theta$ being Hilbert--Schmidt.

We now take up the boundedness-below.  If
$\Theta$ is Hilbert--Schmidt, then $\exp J\Theta$ provides a restricted
symplectomorphism taking $A$ to $\sigma$.  The image of the restricted
symplectomorphism is unitarily implementable, and the quantum
Hamiltonian induced by $\exp t\sigma$ is $-(J\sigma )_a{}^bZ^a\partial
_b$, which is bounded below.  

\end{proof}

This results also implies that the self-adjoint implementation of the
Hamiltonian is essentially unique.

\begin{corollary}
If a classically positive operator $A$ 
with bounded antilinear part is self-adjointly
implementable in the representations determined by $J_1$ and
$J_2$, then its implementations are unitarily equivalent
(modulo an additive constant).
\end{corollary}

\begin{corollary}
A classically positive operator $A$ 
with bounded antilinear part fixes a distinguished
complex structure $J_A$, relative to which $A$ is $J_A$-linear
and $J_A$-anti-self-adjoint.  The operator $A$ is self-adjointly
implementable in the representation determined by a second
complex structure $J'$ if and only if $J_A-J'$ is Hilbert--Schmidt.
\end{corollary}

\section{Connection with Normal Ordering}

As is well-known, even the simplest linear quantum field theories in
Minkow\-ski space contain divergent terms.  For example, the vacuum
energy in a region is $(1/2)\sum \hbar\omega$, where $\omega$ runs over
all the independent modes.  The standard prescription for dealing with
these divergences is {\em normal ordering,} that is, writing all
creation operators before annihilation operators, thus eliminating the
infinite c-number terms.  Of course, normal ordering will not
distinguish between two operators differing by a {\em finite}
c-number, but will reduce them to the same operator.  For this reason,
normal ordering operators may lose certain important physical
information, for example, Casimir-type effects.

In this section, though, we are concerned with a more severe question: 
supposing that one knows that a self-adjoint Hamiltonian exists,
can it necessarily be given by normal-ordering the classical
Hamiltonian (modulo a finite c-number term)?
We shall find that the answer may be No.  
In other words, there are at least in principle linear quantum field
theories which require more than normal ordering to be successfully
renormalized.
This points up the
delicacy of the issues involved in analyzing the quantum Hamiltonian.

While in elementary examples, there is no ambiguity in what is meant by
normal ordering an operator, in the present, very general, context, some
care is needed to make this precise.  This will now be explained.

Let $A$ be a classically positive Hamiltonian, and let $\sigma$ and
$\Theta$ be as in the previous section:  $\sigma$ is a $J$-skew,
$J$-linear, $J$-self-adjoint map and $\Theta$ is a $J$-antilinear,
$J$-symmetric bounded operator with $\sigma =\e ^{-J\Theta}A\e
^{J\Theta}$.  We put $D=-J\sigma$.  Then, using the representation by
creation operators $Z^\alpha$ and destruction operators $\partial
_\alpha$ as described in the previous paper, the quantum Hamiltonian
corresponding to $A$ is the image of $D_\alpha{}^\beta Z^\alpha
\partial _\beta$ under conjugation by the Bogoliubov transformation
induced by $\exp J\Theta$.  Notice that with this definition, the
Hamiltonian has the same spectrum as $D_\alpha{}^\beta Z^\alpha\partial
_\beta$, and in particular has zero as its minimum.

At a formal level, the 
normal ordering is accomplished in the usual way, and one finds
\begin{equation}
  \H =\H _{\rm normal}+E_0
\end{equation}
where the normal-ordered Hamiltonian is
\begin{equation}\label{icarus}
 \H _{\rm normal} ={\overline C}_{\alpha\beta} Z^\alpha Z^\beta
  +B_\alpha{}^\beta Z^\alpha \partial _\beta +C^{\alpha\beta}
   \partial _\alpha \partial _\beta \, ,
\end{equation}
with
\begin{eqnarray}
 {\overline C}&=&J\cosh \Theta D\sinh\Theta\\
 B&=& \cosh\Theta D\cosh\Theta +\sinh\Theta D\sinh\Theta\\
  E_0&=&\trace \sinh\Theta D\sinh\Theta\label{brek}
\end{eqnarray}
in matrix form.

The difficulty in making the relations (\ref{icarus})--(\ref{brek})
precise is not merely in the fact that it is hard to analyze the
individual quantities $B$, $\overline C$, $E_0$.  One has to decide
what sort of properties are required of these in order to say that one
has successfully renormalized the Hamiltonian by normal ordering.  In
the simplest case, one could require that $E_0$ is finite, and that $\H
_{\rm normal}$ be well-defined by term-by-term action on (at least)
a dense family of polynomials.  However, one could also imagine a more
general situation, where the domain consisted of functions $\Psi (Z)$
such that, while the actions of the individual terms in (\ref{icarus})
did not give elements in the Hilbert space, there were nevertheless
cancellations so that the net result was indeed an element of the
Hilbert space.  Indeed, there are even more extreme possibilities.  One
could envisage situations in which $E_0=\pm\infty$, but the elements in
the domain are chosen so that, with a proper limiting procedure, the
quantity $\H |\Psi\rangle$ is well-defined even though 
$\H _{\rm normal}|\Psi\rangle$
is not separately!\footnote{Just this sort of thing would occur if one
defined $\H |\Psi\rangle$ 
by a sequence of formal operations which amounted to
conjugation by the Bogoliubov transformation $\exp J\Theta$.}  Thus at
some point one must decide what sort of regularity the notion of
``normal ordering'' requires; otherwise, saying that normal ordering
suffices to renormalize the Hamiltonian becomes a statement with no
force.  At present, we shall assume it requires $E_0$ to be finite.
This is very weak.

\begin{theorem}\label{albert}
Let $A$ be classically positive with bounded antilinear part.
The minimum of the normal-ordered Hamiltonian is
\[ -E_0=-\trace\sinh\Theta D\sinh\Theta\]
if this is finite.  If it is finite,
it is negative.  If this is infinite, normal ordering
does not suffice to renormalize the Hamiltonian.
\end{theorem}

\begin{proof}
We have mentioned everything except the negativity.  But
$\sinh\Theta D\sinh\Theta$ is a positive symmetric form.  

\end{proof}

Since one can arrange for $\Theta$ to be Hilbert--Schmidt but $E_0$
divergent, this
implies that in principle at least that there are linear
quantum field theories for which self-adjoint Hamiltonians exist, but
they cannot be realized by normal-ordered operators:  some more
sophisticated renormalization is required.  There would be no general
reason for rejecting such Hamiltonians as unphysical, although in a
particular system one might have physical arguments that
normal ordering should suffice to regularize the theory.

For classically positive Hamiltonians, finiteness of the normal-ordered
ground-state energy implies existence of the classical Hamiltonian as a
self-adjoint operator:

\begin{theorem}\label{liese}
Let $A$ be classically positive
with bounded antilinear part, and suppose the normal-ordered ground
state energy $E_0$ is finite.  Then the Hamiltonian is a self-adjoint
operator.
\end{theorem}

\begin{proof}
Let $v_j$ be a $J$-orthonormal basis, and let 
$D=\int _{[\lambda _0,\infty )} 
\lambda \d E(\lambda )$, where $\lambda _0>0$.  Then we are given that
\[ \sum _j \bigl( v_j,\sinh\Theta \int _{[\lambda _0,\infty )}\lambda
  \d E(\lambda ) \sinh\Theta v_j\bigr) \]
converges.  The sum and integration are of non-negative terms, so the
convergence is absolute.  This quantity dominates $\trace \sinh\Theta
\lambda _0I\sinh\Theta$ (where $I$ is the identity).  Since $\lambda
_0>0$, this implies $\sinh\Theta$ is Hilbert--Schmidt, which implies
$\Theta$ is.  

\end{proof}

In general, it is hard to work out $\Theta$ from the normal-ordered
Hamiltonian. In principle, one can work out the ground-state energy
exactly as $(1/4)\trace _\R \bigl( |A| -B\bigr)$, where
$|A|=\sqrt{-A^2}$; this expression does not require a knowledge of
$\Theta$.  However, this quantity is usually too awkward to work with
directly in applications.  It is useful to have some approximate
formulas in terms of $B$ and $C$, which can be worked out directly from
$A$ and $J$.

\begin{theorem}\label{jason}
If the Hamiltonian is classically positive with bounded
antilinear part and the
normal-ordered ground-state energy is finite, it is bounded by
\[ -E_0\geq -\trace CB^{-1}C\]
and also by
\[ -E_0\geq -\trace \sqrt{{\overline C}^T\overline C}\, .\]
If $A$ is classically positive with bounded antilinear part
and either of the quantities on the right is finite,
then the Hamiltonian is self-adjoint and can be renormalized by normal
ordering.
\end{theorem}

\begin{proof}
We have $B\geq \cosh\Theta D\cosh\Theta$ as (densely-defined)
symmetric forms.  Thus $B^{-1}\leq (\cosh\Theta D\cosh\Theta )^{-1}$ 
as symmetric forms.  Thus
\begin{eqnarray*}
  \trace CB^{-1}\overline C
  &\leq&\trace J\sinh\Theta D\cosh\Theta (\cosh\Theta D\cosh\Theta
)^{-1} \cosh\Theta D\sinh\Theta J\\
  &=&-E_0\, .
\end{eqnarray*}

We have
\begin{eqnarray*}
  {\overline C}^T\overline C&=&\sinh\Theta D\cosh ^2\Theta
D\sinh\Theta\\
  &=&\sinh\Theta D^2\sinh\Theta +(\sinh\Theta D\sinh\Theta )^2\, .
\end{eqnarray*}
However, since the first form is positive and symmetric we have
\[ \sqrt{{\overline C}^T\overline C}\geq \sinh\Theta D\sinh\Theta\]
as forms, and so
\[ -E_0\geq\trace \sqrt{{\overline C}^T\overline C}\, .\]
\noindent
\end{proof}
  
\section{Quantum Inequalities}\label{vort}

In this section, I show how to adapt the previous arguments to
a certain important family of situations where $\omega (v,Av)$ is a
positive indefinite form.  I will begin by discussing the class of
operators to be considered and its significance.  Then I shall give the
results.

The proofs of the theorems in this section are fairly lengthy.  This is
ultimately bound up with technical problems at the boundary of the
space--time region whose energy--momentum content is to be measured. 
A brief discussion of this is given at the end of this section, after
theorem~\ref{greenspan}, but some readers may want to look at
this before
the proofs of theorems~\ref{krypton} and \ref{greenspan}.

\subsection{The Sorts of Results Sought}

Ford (1978) was the first to show that
temporal averaging could bound some of the local negative energies
encountered in quantum field theory.
In the case of the Klein--Gordon field
on Minkowski space, he and Roman (1997) proved that
\begin{equation}\label{ratfink}
  \langle\Psi |\int _{-\infty}^\infty {\widehat T}_{00}(t,0,0,0)
   b(t)\, \d t|\Psi\rangle /\langle\Psi |\Psi
  \rangle \geq -{\frac{3}{32\pi ^2}} {\frac{\hbar c}{(ct_0)^4}}\, ,
\end{equation}
where
\begin{equation}
  b(t)={\frac{t_0/\pi}{t^2+t_0^2}}
\end{equation}
is a sampling function of area unity and characteristic scale $\sim
t_0$.
Following Ford, lower bounds on energy operators for relativistic
quantum field theories are known as {\em quantum inequalities}.

In the past years, quantum inequalities of increasing generality have been
established.


\begin{itemize}

\item  
For the massless field in two-dimensional Minkowski space,
there is a broad class of elegant results (Flanagan 1997).  However, the
divergences of this theory are significantly softer than in four
dimensions.  It is especially problematic to draw conclusions about the
boundedness of four-dimensional energies from two-dimensional results.

\item
For the energy density measured by static observers in static
space--times, Pfenning and Ford (1997, 1998) established quantum inequalities
for Lorentzian sampling functions, and Fewster and Teo (1999) for more general
sampling functions.

\item
The results contained in the present papers were released (Helfer 1999a,b).

\item
A general (applicable to a Klein--Gordon 
field in a globally hyperbolic space--time)
quantum inequality for the energy density was given by
Fewster (2000).
Similar results were established for the Dirac field in curved
space--time (Fewster 2002)
and for the spin-one field in curved space--time (Fewster and Pfenning 2003).

\item
Very little is known about bounds on the four-momentum density
of the quantum field.  There is one result, in Minkowski space (Helfer
1998). 

\item
It was shown by Fewster and Roman (2003) that ``null energy inequalities'' do
not exist, that is, that averages of the component ${\hat
T}_{ab}l^al^b$ of the
stress--energy operator 
along a null geodesic with tangent $l^a$ are unbounded below,
even in Minkowski space.

\end{itemize}

One would like to generalize these results to apply
not just to the energy
density, but to other components of the stress--energy tensor.  This is
because classical matter fields satisfy not only the Weak Energy
Condition (which requires the energy density to be positive) but also
the Dominant Energy Condition (which requires the four-momentum
density to be future-pointing).  One would like to know what bounds
there are on violations of the Dominant Energy Condition.

I think that the approach used by Fewster and co-workers
can be extended to give such results.  I also believe that that approach gives a
very useful ``hands-on'' understanding of the quantum inequalities in terms of
the geometry of space--time as reflected in the ultraviolet Hadamard
asymptotics.  On the other hand, the present paper gives a rather different
perspective.  

The previous two sections can be interpreted as showing that classical
positivity enforces boundedness-below of the quantum Hamiltonian.  This is very
close to saying that classically positive measures of energy (including
classically positive averages of the stress--energy) should have
quantum-inequality counterparts; the difference lies in technicalities
associated with the finite extent in space--time of the regions over which the
averages are taken.  Leaving aside these technicalities for the moment, then, we
have a suggestive
argument for a very broad class of quantum inequalities which emphasizes
the general algebraic properties of the classical measures of energy rather than
the ultraviolet asymptotics of the quantum field theory.  (The ultraviolet
asymptotics do play an important if brief role in showing that $A_-$ is
bounded.)

For generic space--times, there is no preferred vacuum state and no
preferred associated quantization.  Rather, one has a family of
unitarily equivalent (modulo infrared issues) Hadamard quantizations.
Different choices of such quantization lead to different
normal-ordering prescriptions and normal-ordered Hamiltonians which
differ by c-numbers.  This is bound up with the well-known
ambiguities in fixing the c-number part of the stress--energy operator
in generic space--times.  Thus, a specific numerical lower bound on the
Hamiltonian is only meaningful given choices which resolve these
ambiguities.  Given our current lack of understanding of how to effect
these resolutions, such numerical values would be data of no clear
significance.

There are, however, two sorts of results which would be of immediate
significance.  One of these would be asymptotic formula for lower
bounds, as the sampling function becomes more and more localized.  (For
example, as $t_0\downarrow 0$ in (\ref{ratfink}).)  In such cases,
because the energy densities diverge, the c-number ambiguities in the
stress--energy become insignificant.  While I believe that such results
can be derived using the techniques of this paper, they require lengthy
computations which are too far out of the main line of argument.  I
shall give such results elsewhere.

The second sort of presently-useful result would be a general proof
that temporally-averaged Hamiltonians are bounded below.  This is a
statement which is meaningful even in the face of the c-number
ambiguities, and it is this result which will be proved in the remainder
of this section.

In order to treat such very general cases, it is probably necessary to
pass to compactly supported sampling functions.  (For otherwise, with
no assumptions about the space--time, one has no control about how
small data from initially spatially distant regions propagate inwards
and are amplified by the space--time geometry.)  I shall not use the
stress--energy localized to a world-line, but rather to a compact
four-volume.  Probably one can obtain parallel results for
world-lines, but the proofs would be longer.  For four-volumes, we
can appeal to very general results in distribution theory (cf. the
proof of Proposition~\ref{pincers}, below).

\subsection{Structure of the Hamiltonians; Boundedness-Below}

Let $(M,g_{ab})$ be an oriented, time-oriented space--time, globally
hyperbolic with compact Cauchy surfaces.  (The restriction to compact
Cauchy surfaces is a technical device to simplify the analysis and
remove infrared ambiguities.  It is not physically significant.
The analysis is all local, and the spatial dimensions can be arbitrarily
large.)
Consider the quantum theory
of a real scalar field governed by the equation
\begin{equation}
  (\nabla ^a\nabla _a +m^2)\phi =0\, ,
\end{equation}
where $m^2\geq 0$.  The associated classical stress--energy is
\begin{equation}
  T_{ab}=\nabla _a\phi \nabla _b\phi -(1/2)g_{ab}(\nabla _c\phi
  \nabla ^c\phi -m^2\phi ^2)\, .
\end{equation}
We shall say a smooth symmetric compactly-supported tensor field
$f^{ab}$ is {\em future-directed} if it is at any point a sum of
terms $t^au^b+u^at^b$ with $t^a$, $u^a$ future-pointing.  The content
of the classical {\em dominant energy condition} 
(which holds for this classical field) is that $\int
f^{ab}T_{ab}\d {\rm vol}\geq 0$ for any future-directed $f^{ab}$.

Letting $A$ be the generator corresponding to this Hamiltonian, we have
\begin{equation}
  (1/2)\omega (\phi ,A\phi )=\int f^{ab}T_{ab}\d {\rm vol}\, .
\end{equation}
The operator $A$ is not in general classically positive.  In the first
place, the test function may be supported in an arbitrarily small
volume, and so $A$ may have a large kernel.  A more severe problem is
that the smooth fall-off of the test function generally 
leads to a spectrum including points arbitrarily
close to zero.  However, a good fraction of the structure deduced for
classically positive Hamiltonians still applies.

It is appropriate to review what the correspondence between the rather general
approach adopted so far in this paper and the specifics of the present scalar
field are.  

The space $H$ consists of classical solutions to the field equation
whose restrictions to any Cauchy surface have Sobolev regularity $1/2$.  (This
choice of regularity makes all integrals for the symplectic form and the
inner product converge as they should.)  The symplectic form is given by
\begin{equation}
  \omega (\phi ,\psi )=\int _\Sigma \left( \psi {}^*d\phi -\phi {}^*d\psi\right)
\end{equation}
where $\Sigma$ is any Cauchy surface.  It is independent of the surface chosen
by virtue of the field equation.

It should be emphasized that, while our ultimate interest is in quantum fields,
the formulas given so far in this subsection are entirely classical. (Thus $A$ is an
operator on the classical space of solutions --- a Hamiltonian vector
field ---, not a quantum operator.)
The
connection with quantum field theory is made through the specification of a
complex structure $J$, or equivalently, a two-point function.  Let us review
how this comes about.

As mentioned above, there is in general no canonical vacuum state.  However,
there is a class of states which is preferred, the {\em Hadamard} states,
those whose highest-order ultraviolet asymptotics agree with those of the Fock
representation in Minkowski space.
Any one of these can be used as a ``vacuum'' for the construction of the field
theory, by following the usual mathematical pattern of the Fock construction. 
To see this, note that
\begin{equation}
\langle 0|{\hat\phi} (x){\hat\phi}(y)|0\rangle
  =\langle 0|{\hat\phi}^+(x){\hat\phi}^-(y)|0\rangle
\end{equation}
in the usual construction.  On the other hand, if one is given the two-point
function, one can take this equation as {\em defining} the positive- and
negative-``frequency'' projections of the field, and hence the annihilation and
creation operators.  (The term ``frequency'' is used here only to bring out the
correspondence with the usual Fock construction.  ``Frequency'' here is not
the Fourier transform with respect to any
obvious time variable.)  Representations constructed from different Hadamard
states will be unitarily equivalent.

Thus specification of a Hadamard state allows one to determine a mathematical
structure analogous to the usual Fock one, the key portion being a mathematical
analog of the decomposition into positive and negative ``frequencies,'' which allows
the definition of annihilation and creation operators.  Contrariwise, such a
positive-/negative-frequency decomposition of the field operators would
determine a Fock-type quantization.  

The decomposition into positive and negative ``frequencies'' is naturally
encoded in a complex structure $J$, setting $J\phi =\ix (\phi _+ -\phi _-)$. 
Thus the quantization is specified by the complex structure $J$.  In the case of
Hadamard states, it turns our that $J$ can be expressed as a pseudodifferential
operator (cf. section 6 of paper I).

\begin{proposition}\label{pincers}
Let $A$ be the generator associated to a smooth compactly-supported
test field (not necessarily future-directed).  Then $A_-$,
the $J$-anti\-linear
part of $A$, is represented on initial data by an operator with smooth
kernel.  In particular, the operator $A_-$ is compact.
\end{proposition}

\begin{proof}
We make use of the microlocal properties of the two-point functions, 
discovered by Radzikowski (1992) and Junker (1995).  A summary adequate
for understanding this proof is in the paper of Brunetti et al. (1996). 
For the general theory of wave-front sets, see H\"ormander (1983).

The antilinear part of $A$ is got by projecting its positive-to-negative and
negative-to-positive ``frequency'' parts using the two point function; it
corresponds to the two-point kernel
\begin{eqnarray*}
\lefteqn{A_-(y,z) =\int f^{ab}(x)
  \bigl( (\nabla _{x^a}K(x,y))(\nabla _{x^b}K(x,z)) }\\
  &&-(1/2)g_{ab}(\nabla _{x^c}K(x,y))(\nabla ^{x^c}K(x,z))\\
  &&+(1/2)g_{ab}K(x,y)K(x,z)\, \d {\rm vol}(z)\, .
\end{eqnarray*}
Here $K(x,y)$ is the two--point function.  The wave--front set of
$K(x,y)$ is 
\[ {\rm WF}(K)=\{ (x,k;y,l)\mid (x,k)\sim (y,l)\,
 ,\ k\mbox{ is future-pointing}\, \}
.\]
Here and in what follows, it is understood that $(x,k),(y,l)\in T^*M-\{
0\}$; and
$(x,k)\sim (y,l)$ iff there is a null geodesic from $x$ to $y$
with covector $k$ at $x$ and $l$ at $y$.  Thus
\begin{eqnarray*}
{\rm WF} (K(x,y)K(x,z))&\subset &
  \{ (x,k,y,l,z,m)
  \mid (x,k)\sim (y,l)\mbox{ or } (x,k)\sim (z,m)\}\\
  &\cup&\{ (x,k,y,l,z,m)\mid
    (x,k_1)\sim (y,l)\mbox{ and } (x,k_2)\sim (z,m)\\
  &&\quad\mbox{ for some }
  k_1,k_2\mbox{ with } k=k_1+k_2\}\, .
\end{eqnarray*}
When this is integrated against $f^{ab}(x)$ to form $A_-(y,z)$, the
result has
\[ {\rm WF}\, (A_-(y,z))\subset
  \{ (y,l,z,m)\mid (x,0,y,l,z,m)\in{\rm WF}(K(x,y)K(x,z))\mbox{ for some }x\} \, ,
\]
which is empty.  Since the two-point function has a smooth kernel, so
does its restriction to act on initial data.

\end{proof}

\begin{proposition}
Let $A$ be the generator associated to smooth compactly-supported
future directed test field.  Then the spectrum of $A$ lies on the
imaginary axis.
\end{proposition}

\begin{proof}
Consider $A=A_++A_-$ as a perturbation of its $J$-linear part.  The
term $A_+$ is $J$-real anti-self-adjoint and so has purely imaginary
spectrum.  On the other hand, the compactness of $A_-$ means that any
element of ${\rm spec}\, A -{\rm spec}\, A_+$ must be an eigenvalue.
(To see this, suppose one has $\lambda\in {\rm spec A} -{\rm spec A_+}$
and a sequence $v_n$ of unit vectors in the domain of $A$ with
$(\lambda -A )v_n\to 0$.  Multiplying by $(\lambda -A_+)^{-1}$ we have
$v_n-(\lambda -A_+)^{-1}A_- v_n\to 0$.  Since $(\lambda -A_+)^{-1}A_-$
is compact, this implies $(\lambda -A_+)^{-1}A_-$ has an eigenvalue
unity.  However, this implies $\lambda$ is an eigenvalue of $A$.)
However, then the positivity of $A$ implies, as in the proof of
Proposition~\ref{phyllis}, that any such eigenvalue is imaginary.

\end{proof}

\begin{proposition}
Let $A$ be the generator associated to smooth compactly-supported
future-directed test field.  
Then for any $t\in\R$, the spectrum of $g(t)=\exp tA$
lies on the unit circle.
\end{proposition}

\begin{proof}
We consider $g(t)$ as a perturbation of $\exp tA_+$; the latter is
$J$--unitary and so has spectrum on the unit circle.  It follows from
standard perturbation theory (Dunford and Schwartz 1988, Theorem
VIII.1.19) that $g(t)-\exp tA_+$ is compact.  Applying the argument of
the previous proof, we see that any element of the spectrum of $g(t)$
not already in the spectrum of $\exp tA_+$ must be an eigenvalue; but
then that $g(t)$ be a symplectomorphism implies that eigenvalue lies on
the unit circle.

\end{proof}

\begin{theorem}\label{krypton}
Let $A$ be the generator associated to smooth compactly-support\-ed
future-directed test field.  Then there is a (strong)
projection-valued
distribution $\d E(l)$ which provides a spectral resolution of $A$ in the sense
of Fourier transforms:
\[ \int _{-\infty}^\infty \phi (l)\, dE(l)
  =\int _{-\infty}^\infty g(t)\tilde\phi (t)\, dt\]
strongly,
for suitable test functions $\phi$.  As $\phi (l)$ approaches $l$ in a suitable
sense, we have $A =\int \ix l\d E(l)$ strongly on $D(A)$.

The distribution $\d E(l)$ is locally a measure, and this measure is
locally
integrable except possibly at zero; the quantity $l\d E(l)$ is
integrable at zero.  (Here and throughout, these statements are to be
understood strongly, that is, the operators applied to elements of
$H$.)
\end{theorem}

\begin{proof}
This proof is somewhat technical, and makes use of the theory of Fourier
hyperfunctions.  This is the class of generalized functions ${\cal F}'$ 
dual to ${\cal F}
=\{ \phi\mid \phi\, ,{\tilde\phi}$ are both smooth of exponential decay
$\}$.  (Here $\tilde\phi$ denotes the Fourier transform of $\phi$.)
The results we use are contained in the papers of 
Chung and Kim (1995) and Chung et al. (1994).  The topology on
${\cal F}$ is given by the family
of seminorms $\| \phi \| _{k,h} =\sup _{x,n} |\phi ^{(n)}(x)|e^{k|x|}/(h^n n!)$.
A sequence $\phi _j\to 0$ in ${\cal F}$ if for some $k,h>0$ one has $\| \phi _j\|
_{k,h}\to 0$ as $j\to\infty$.

Put
$\lambda =\ix l$, where $l$ is a real parameter (which may acquire a
small imaginary part).  Let $[(\ix l-A)^{-1}]$ denote the jump in $(\ix
l-A)^{-1}$ from above the real $l$-axis to below; this jump is by
definition a hyperfunction. (One defines its pairing with a test function by
integrating slightly above and slightly below the axis; cf. Cerezo et al. 1975.)
We shall interpret this hyperfunction in
the strong sense, that is, as applied to any vector in $H$.  All
integrals of operators in what follows are also to be interpreted in
the strong sense.

We now show that
$[(\ix l-A)^{-1}]$ is not just a hyperfunction, but in fact a Fourier
hyperfunction.  This means that for any $\phi\in{\cal F}
=\{ \phi\mid \phi\, ,{\tilde\phi}$ are both smooth of exponential decay
$\}$, the integral
$\int _{-\infty}^\infty [(\ix l-A)^{-1}]\, \phi (l)\, \d l$ is defined
and varies continuously with $\phi$. 
We have
\begin{eqnarray*}
 \int _{-\infty}^\infty [(\ix l-A)^{-1}]\, \phi (l)\, \d l
  &=&\int _{-\infty}^\infty \left[ (\ix l+\epsilon -A)^{-1}
            -(\ix l-\epsilon -A)^{-1}\right] \, \phi (l)\, \d l\\
  &=&\int _{-\infty}^\infty \left[ \int _0^\infty g(t) 
   \e ^{-t(\epsilon +\ix
l)}\d t \right.\\
  &&\qquad \left. +\int _{-\infty}^0 g(t)\e ^{t(\epsilon -\ix l)}\d t
            \right] \, \phi (l)\, \d l\\
  &=&\int _{-\infty}^\infty  \int _{-\infty}^\infty g(t)
    \e ^{-\epsilon |t| -\ix lt}\d t
            \, \phi (l)\, \d l\\
  &=&\int _{-\infty}^\infty  g(t)\e ^{-\epsilon |t|}{\tilde\phi}(t)\, \d
t\\
  &=&\int _{-\infty}^\infty  g(t){\tilde\phi}(t)\, \d t\, ,
\end{eqnarray*}
the limit $\epsilon\downarrow 0$ being understood.
This is well-defined, since the class ${\cal F}$ is invariant under
Fourier transform.  To see that it depends continuously on $\phi$, we
note that if $\phi _j$ is a sequence tending to zero in ${\cal F}$, then
there exists $h>0$ such that $\sup {\tilde\phi}_j(t)\exp h|t|$ tends to
zero as $j\to\infty$.  
However, since $\| g(t)\| _{\rm op}\leq M\exp h|t|/2$
(say --- using the Hille--Yoshida--Phillips
estimate and the fact that the spectrum of $A$ is purely
imaginary), the integrals tend to zero as $j\to\infty$.

Some comments are in order at this point:  (a) We have just shown that
the jump in the resolvent is the Fourier transform of $g(t)$.  (b) The
integral displayed above must lie in the domain of $A$ (since the class
${\cal F}$ is invariant under differentiation).  (c) These results are
independent of positivity properties of $A$.  In general, then, groups
of type zero (in the semigroup sense) have generators which can be
analyzed in terms of Fourier hyperfunctions.  These
generalized functions admit a useful microlocalization.

A direct argument to establish positivity is probably possible, but
partly for technical reasons and partly for its future utility, I shall
give another.  Every Fourier hyperfunction can be viewed as an
initial datum for the heat equation.  If the corresponding solution for
$t>0$ is everywhere non-negative, then the hyperfunction is a measure.  
This result is due to Chung et al. (1995), and can be thought of as an
extension of the Bochner--Schwartz theorem.
(In
this context, the initial datum is called an Aronszajn trace.)

The solution to the heat equation with initial value $[(\ix l-A)^{-1}]$
is
\begin{eqnarray*}
 U(x,t)&=&(4\pi t)^{-1/2}\int _{-\infty}^\infty \e ^{-(x-l)^2/(4t)}
  [(\ix l-A)^{-1}]\, \d l\\
  &=&2^{-1/2}\int _{-\infty}^\infty g(s)\e ^{-\ix xs-ts^2}\, \d s\, .
\end{eqnarray*}
For $t>0$, this maps to the domain of $A$.  We also note the identities
\[ U(x,t)U(x,t)=(\pi /4t)^{1/2}U(x,t/2)\]
and
\[ \overline{U(x,t)} =
  2^{-1/2}\int _{-\infty}^\infty g(-s)\e ^{-\ix xs-ts^2}\, \d s\, ,\]
where the overline indicates the complex conjugate.  Using these, we
have
\begin{eqnarray*}
 \omega\bigl( v,AU(x,t)v\bigr)
  &=&(4t /\pi )^{1/2}\,\omega\bigl( v,AU(x,2t)U(x,2t)v\bigr)\\
  &=&(4t/\pi )^{1/2}\, \omega\bigl( \overline{U(x,2t)}v,AU(x,2t)v\bigr)
\, ,
\end{eqnarray*}
and this is positive.

We know at this point that for any $v$, the quantity 
$\omega (v,A[(\ix l-A)^{-1}]v )$ is an exponentially tempered measure. 
It is easily verified that the form $\omega (\cdot ,A[(\ix l-A)^{-1}]\cdot )$ 
is Hermitian, so by polarization $A[(\ix l-A)^{-1}]$ is an
operator-valued exponentially tempered measure (in the strong sense).
If we could divide this by $\ix l$, we could conclude that $[(\ix
l-A)^{-1}]$ is a measure.  However, this is not obviously possible.

For any test function $\phi (l)\in {\cal F}$, write $\phi (l) =\phi
(0)e^{-l^2}+(\phi (l)-\phi (0)e^{-l^2})$.  Then $(\phi (l)-\phi
(0)e^{-l^2})$ is $l$ times a smooth function.  Using this decomposition,
it is easy to see that $[(\ix l-A)^{-1}]$ extends to a linear form on
those continuous functions of exponential decay which are $C^1$ at
the origin.  In particular, the Fourier hyperfunction $[(\ix l-A)^{-1}]$
is a distribution.  Since $l[(\ix l-A)^{-1}]$ is a measure, we must have
$[(\ix l-A)^{-1}] =\alpha \delta '(l) +\mu '$ for some $\alpha$ and some
measure $\mu '$.  (We are following the convention where distributions
are represented by ``generalized functions'' under the integral sign,
and so the measure is represented as $\mu '$ or $\mu '\d l$.  This is
only a symbolic representation, and is not meant to assert any
regularity of the measure with respect to Lebesgue measure.
The prime is present so as to have the formal correspondence $d\mu =\mu 'dl$, but we
we are not interested in trying to construct $\mu$ or giving $\mu '$ meaning oas a
derivative.
We should
more carefully absorb the $\d l$'s into $[(\ix l-A)^{-1}]$ and $\delta
'(l)$ and write simply $\d\mu (l)$.)  We caution that $\mu
'\d l$ is known to be locally finite only on $\R -\{ 0\}$.

The coefficient $\alpha$ must vanish.  To see this, note that we have
(strongly)
\[ \alpha =-\lim _{a\downarrow 0}\int _{-\infty}^\infty 
   [(\ix l-A)^{-1}] l\e ^{-l^2/(2a^2)}\d l\, \]
and hence
\begin{eqnarray*}
 A\alpha &=&-A\lim _{a\downarrow 0}\int _{-\infty}^\infty 
   [(\ix l-A)^{-1}] l\e ^{-l^2/(2a^2)}\d l\\
  &=&-\lim _{a\downarrow 0}\int _{-\infty}^\infty 
   \ix [(\ix l-A)^{-1}] l^2\e ^{-l^2/(2a^2)}\d l\\
  &=&0\, .
\end{eqnarray*}
Passage from the first to the second line is justified by use of the
Fourier transform for $[(\ix l-A)^{-1}]$, or by the identity
$A(\ix l-A)^{-1}=-1 +\ix l(\ix l-A)^{-1}$; passage from the second to the
last by the fact that $l[(\ix l-A)^{-1}]$ is a measure.  Now let $v\in H$
and let $w=\alpha v$.  It is easy to see that as a distribution 
in space--time,
the quantity $w$ vanishes on $M-{\rm supp}\, f^{ab}$.  On the other
hand, since $Aw=0$, the local positivity of the form $T_{ab}f^{ab}$
implies that $w$ vanishes on the interior of ${\rm supp}\, f^{ab}$.
(To see this, note that we may replace $[(\ix l-A)^{-1}] l$ in the integrand
with $[(\ix -A)^{-1}] (\ix A)$, which will annihilate anything supported outside
of the support of $f^{ab}$.)
Thus if $w$ were known to be smooth, we would have $w=0$.  However, it
is easy to check that for any $u\in H$ we have $\omega (u,w)=-\omega
(\alpha u, v)$.  By the arguments just given, this vanishes for smooth
$u$; since the smooth $u$'s are dense in $H$, it vanishes always and $w=0$.
Hence $\alpha =0$.

We shall now write $\d E(l)=[(\ix l-A)^{-1}]\d l$.  It is a
projection-valued distribution which is locally a projection-valued
measure.  To see that it is
projection-valued, note that for any test function $\phi (l)$ we have
\begin{eqnarray*}
 \left( \int _{-\infty}^{\infty} [(\ix l-A)^{-1}]\, \phi (l)\, \d l
    \right) ^2
  &=&\int _{-\infty}^\infty g(t)g(s){\tilde\phi}(t){\tilde\phi}(s)\, 
    \d t\, \d s\\
  &=&\int _{-\infty}^\infty g(u){\tilde\phi}(u-s){\tilde\phi}(s)
  \d u\, \d s\\
  &=&\int _{-\infty}^{\infty} [(\ix l-A)^{-1}]\, \phi (l)^2\, \d l\, .
\end{eqnarray*}

To establish the compatibility of the
decompositions with the symplectic structure, notice that
\begin{eqnarray*}
  \omega (\int \phi (l)\d E(l) v,\int \psi (k)\d E(k) w)
  &=&\omega (v,\int\phi (-l)\d E(l)\int\psi (k)\d E(k) w)\\
  &=&\omega (v,\int \phi (-l)\psi (l)\d E(l)w)\, .
\end{eqnarray*}
For a subspace $H_S=\int _S\d E (l)H$ to be real, the set $S$ must be
symmetric (up to terms of $\d E$-measure zero).
For symmetric sets, the equation above shows that $H_S\cap
H_{S'}=\{ 0\}$ if $S\cap S'=\emptyset$.  Thus the spectral decomposition
by $\d E$ respects the symplectic structure.  That $\omega$ must be
strongly non-degenerate on each $H_S$ follows.

To establish $A=\int\ix l dE(L)$ strongly on $D(A)$, we consider
\[ \int _{-\infty}^\infty [(\ix l-A)^{-1}] \phi (l)\, dl
  =\int _{-\infty}^\infty g(t)\tilde\phi (t)\, dt\, .\]
Take, for instance, the function $\phi =l\exp -\epsilon l^2/2$ and $v\in D(A)$. 
Then
\[ \tilde\phi = -(2\pi \epsilon )^{-1/2}
 \frac{d}{dt} e^{-t^2/(2\epsilon )}\, du\, .\]
Substituting this into the previously displayed equation and integrating by
parts, we get
\[ \int _{-\infty}^\infty [(\ix l-A)^{-1}] e^{-\epsilon l^2/2}\, dl v
  =(2\pi \epsilon )^{-1/2} \int _{-\infty}^\infty
     Ag(t)e^{-t^2/(2\epsilon )}\, dt\, .\]
Taking $\epsilon\downarrow 0$ gives the required result.

\end{proof}

It is natural to wonder about the integrability of $dE(l)$ at infinity.  This is
a delicate issue, on account of the non-scalar nature of the measure.  I believe
it is possible to get fairly general results, but I shall not attempt these
here.  We saw above that we can get useful specific results
by exploiting the Fourier transform relation
\begin{equation}
\int _{-\infty}^\infty [(\ix l-A)^{-1}] \phi (l)\, dl
  =\int _{-\infty}^\infty g(t)\tilde\phi (t)\, dt
\end{equation}
for specific suitable functions $\phi (l)$.  In particular, if $\phi$ is a
constant (or exponential of pure frequency), then $\tilde\phi$ is a
delta-function and the above equation defines the left-hand side strongly. 
Similarly, for $\phi (l)=le^{\ix kl}$ the Fourier transform exists as the
derivative of a delta-function, and the equation defines the left-hand side
strongly on $D(A)$.  Thus the class of $\phi$'s for which $\int _{-\infty} [(\ix
l-A)^{-1}]\phi (l)\, dl$ exists strongly may be extend to include (for example)
those which are sums of constants plus continuous compactly supported functions;
and the integral may be defined strongly on $D(A)$ for sums of linear functions
plus continuous compactly supported functions.

The delicate issues involved in developing a very general theory of this have to
do with clarifying precisely the set of admissible $\phi$'s and its topology. 
However, the observations we have made will be enough for this
paper.

While we are not guaranteed the sort of canonical form we had for
classically positive operators, we may still draw some conclusions by
considering $H$ as a
limit of spaces.

\begin{proposition} 
Let $A$ be the generator associated to smooth compactly-supported
future directed test field.  Then the associated normally-ordered
Ham\-iltonian operator is bounded below if $A_-$ is trace-class; in this
case, there is a bound 
\[ -E_0\geq -{\rm tr}\sqrt{ {\overline C}^T\overline{C} }\, .\]
\end{proposition}

\begin{proof}
In order to make use of the results of previous sections on classically positive
Hamiltonians, we shall introduce a modified family of operators $A_\epsilon$
which are classically positive and tend to $A$ as $\epsilon\downarrow 0$.

Let $b(l)$ be a continuous bump function, supported on $[-1,1]$, identically
unity in a neighborhood of the origin, and symmetric.  Let $A_\epsilon =\int _\R
\ix l (1-b(l/\epsilon ))\, dE(l)$.  We may regard $A_\epsilon$ either as an
operator on $H$, or, when convenient, on the subspace $H(\epsilon )
\int _\R (1-b(l/\epsilon
))\, dE(l) H$.  The discussion of the compatibility between $dE(l)$ and $\omega$
at the end of the last proof shows that $\omega$ restricts to be non-degenerate
on $H(\epsilon )$, and then $A_\epsilon$ is classically positive on $(H(\epsilon
),\omega )$.
We have $A_\epsilon v\to Av$ as $\epsilon \downarrow 0$
for $v\in D(A)$, since $l\d E(l) v$ is a measure (and the mass of $\{
0\}$, that is, the coefficient $\alpha$ in the previous proof, is zero).

Now let $|\Psi \rangle$ be any Hadamard state of norm
unity.  This means that in the
holomorphic representation $\Psi (Z)$ is a polynomial whose coefficients
are represented by smooth fields on space--time; in particular, these
coefficients lie in (tensor products of) $D(A)$.  Thus we may compute
\[ \langle \Psi |\H |\Psi\rangle =\lim _{\epsilon\downarrow 0}
    \langle\Psi |{\H}_\epsilon |\Psi\rangle\, ,\]
where ${\H}_\epsilon$ is the Hamiltonian defined from $A_\epsilon$
by normal ordering.  (Brunetti et al. 1996 showed that $\H$
may be defined by normal ordering.)  But
we know a lower bound of $\H _\epsilon$ is $-\tr |(A_\epsilon )_-|$.
Now in fact the lower bounds are monotonically decreasing with
$\epsilon$.  This follows from the fact that for any fixed $\epsilon
_0>0$, a fixed $\Theta$ can be found which simultaneously provides a
similarity of all $A_\epsilon$ with $\epsilon >\epsilon _0$ to
generators of orthogonal groups.  It follows that
\[ \inf\H \geq\lim _{\epsilon\downarrow 0}\inf {\H}_\epsilon\, ,\]
where $\inf{\H}_\epsilon$ denotes the infimum of the spectrum.

Similarly, it follows from the formula for ${\overline C}^TC$ in
theorem~\ref{jason} that ${\overline C}^T_\epsilon C_\epsilon$ is a family
of symmetric positive forms, which are (as forms) increasing as
$\epsilon \downarrow 0$.  But as we know that $C_\epsilon\to C$
strongly, we have
\[ \lim _{\epsilon\downarrow 0}\tr 
  \sqrt{{\overline C}^T_\epsilon C_\epsilon }\leq \tr
  \sqrt{{\overline C}^T C} \, .\]

\end{proof}

We are now in a position to establish
the existence of a very large class of quantum inequalities.

\begin{theorem}\label{greenspan}
Let $A$ be the generator associated to smooth compactly-support\-ed
future--directed test field.  Then the associated quantum Hamiltonian is
bound\-ed below.
\end{theorem}

\begin{proof}
It only remains to note that since $A_-$ has a smooth kernel, it is
trace-class.  (See e.g. Treves 1980.)

\end{proof}

The argument for this result has been very technical, and I wish to
comment here on why this is.

First, it must be emphasized that because in general the operators we
are dealing with are not self-adjoint (nor unitary), merely having some
control over their spectrum tells us very little.  For example, suppose
we have an operator with a discrete set of eigenvalues:
\begin{equation}
 \sum _j\lambda _jE_j\, ,
\end{equation}
where the $E_j$'s are projections.  If the operator is self-adjoint,
then the $E_j$'s are {\em orthogonal } 
projections, and in particular, {\em
uniformly bounded as operators}.  However, in the more general case,
the $E_j$'s may have diverging bounds.  (That is, we may be able to find
a unit vector $v_j$ so that $\| E_jv_j\|\to\infty$.)  Thus it is quite
possible to have $\lambda _j\to 0$ but still have the sum above
represent an {\em unbounded } operator, or to have the bounds of
$\lambda _jE_j$ not tend to zero.

Just these sorts of concerns are present in the regime $l\approx
0$ for the operator $A$.  This can be understood by considering its
interpretation in space--time, as follows.  Since $l$ is the Fourier
transform variable to $t$, we may expect that the behavior of $A$ near
the spectral parameter $l=0$ is related to the $t\to\pm \infty$
asymptotics of $g(t)$.  In space--time, this corresponds to flowing
along the Hamiltonian vector field determined by $\int f^{ab}T_{ab}\d
{\rm vol}$ for very long times.  Now, if we start with some general
solution $\phi$ and flow along this vector field, whatever oscillations
$\phi$ has within the region $f^{ab}\not= 0$ will tend to pile up on the
future and past boundaries of that region.  Thus as $t\to\pm\infty$, the
quantity $g(t)\phi$ will be approximately some average value in the
interior of $f^{ab}\not= 0$, but quite scrunched up near the boundary. 
It is very possible that this results in $\d E(l)$ not being integrable
at $l=0$.

A second difficulty is that we do not have very good control over the
quantization of $A$, compared to that for classically positive
operators.  We know, from the work of Brunetti et al., that $A$ {\em is
} self-adjointly implementable by normal ordering, but we do not have
the sorts of explicit control over its lower bound that we had in the
previous section.  This is related to the first difficulty, in that what
prevents us from having this control is the fact that the operator
$\Theta$ may not be bounded, which is again due to the $l\approx 0$
behavior of $A$.

To get around this lack of control of the quantization, we approximated
the operator $A$ by operators $A_\epsilon$.  This approximation, though,
was rather weak, necessitating some further, indirect steps.

From a physical point of view, the
differences between $A$ and the limit of the $A_\epsilon$'s,
and $\H$ and the $\H _\epsilon$'s, are measures
of the importance of effects the boundary of the region
$f^{ab}\not= 0$, which one would hope are unimportant.  After all, the
point of having $f^{ab}$ approach zero smoothly at the boundaries was
precisely to try to minimize edge effects.  However, at least the
present argument does not show, for example, that $\lim
_{\epsilon\downarrow 0}\inf\H _\epsilon =\inf\H$.

\section{Summary and Conclusions}

These papers were motivated by the desire to understand some surprising
and at least apparently pathological results for quantum fields in
curved space--time.  The worst of these is that, in generic
circumstances, the Hamiltonians are unbounded below.  This is absolutely
counter to one's expectations.  If in fact these field theories do
describe the real world, then one must explain why these pervasive
arbitrarily negative energies do not lead to instabilities.

As emphasized in the introduction to Paper I, the present analysis is
only a step to understanding these properties.  We have aimed here to
get a clear statement of what the mathematical structure of the theory
actually is.  We have seen that the assumption that the classical
energy function of a Hamiltonian system is strictly positive provides a
very strong restriction on its structure, somewhat analogous to the
compactness of the energy surfaces in the finite-dimensional case.
This gives one good mathematical control, and one can say under what
conditions a self-adjoint quantum Hamiltonian exists.  We have seen
that there is an intimate connection between self-adjointness (or, at
the level of finite evolutions, unitarity), and boundedness below.  All
self-adjoint quantizations are unitarily equivalent (modulo additive
constants), and all are bounded below.

These positive mathematical results throw the pathological features
into stronger relief.  In generic circumstances, the Hamiltonians for
temporal evolution of quantum field theories are {\em neither }
self-adjoint {\em nor } bounded below.  Typically, it is only
temporally-averaged energy operators which are bounded below (and are
self-adjoint):  this is the force of the quantum inequalities, proved
in the previous section.

The resolution of the pathologies will require physical input.  I
have shown earlier that, at least in many circumstances, there are
limits from quantum measurement theory on the detection of negative
energy densities (Helfer 1998).  However, this is at present far from an
explanation of why the predictedly generic arbitrarily negative energy
densities seem to have no role in the world.

\paragraph{Acknowledgments}
It is a pleasure to thank members of the
University of Missouri--Columbia Mathematics Department, 
especially Peter Casazza, Nigel Kalton, Yuri
Latushkin and Stephen Mont\-gomery--Smith, for helpful
conversations.

\end{document}